# Image quality enhancement in wireless capsule endoscopy with Adaptive Fraction Gamma Transformation and Unsharp Masking filter

[1]Rezvan Ezatian[*], [1]Donya Khaledyan, [1]Kian Jafari, [2]Morteza Heidari,
[3]Abolfazl Zargari Khuzani, [4]Najmeh Mashhadi
[1]Faculty of Electrical Engineering, Shahid Beheshti University, Tehran, Iran.
[2]School of Electrical & Computer Engineering, University of Oklahoma, Norman, USA
[3]The Department of Electrical and Computer Engineering, University of California, Santa Cruz, USA
[4]The Department of Computer Science and Engineering, University of California, Santa Cruz, USA
* r.ezatian@mail.sbu.ac.ir

*Abstract*— **Wireless Capsule Endoscopy (WCE) presented in 2001 as one of the key approaches to observe the entire gastrointestinal (GI) tract, generally the small bowels. It has been used to detect diseases in the gastrointestinal tract. Endoscopic image analysis is still a required field with many open problems. The quality of many images it produced is rather unacceptable due to the nature of this imaging system, which causes some issues to prognosticate by physicians and computer-aided diagnosis. In this paper, a novel technique is proposed to improve the quality of images captured by the WCE. More specifically, it enhanced the brightness, contrast, and preserve the color information while reducing its computational complexity. Furthermore, the experimental results of PSNR and SSIM confirm that the error rate in this method is near to the ground and negligible. Moreover, the proposed method improves intensity restricted average local entropy (IRMLE) by 22%, color enhancement factor (CEF) by 10%, and can keep the lightness of image effectively. The performances of our method have better visual quality and objective assessments in compare to the state-of-art methods.**

*Keywords— wireless capsule endoscopy, image enhancement, adaptive fraction gamma transformation, Unsharp Masking filter, color restoration.*

## I. Introduction

Diseases related to the gastrointestinal (GI) tract seriously threaten human health nowadays. With early diagnosis, it is possible to stop manifold infections in the GI tract. Wireless Capsule Endoscopy (WCE) is growing to a technology revolution that lets the visualization of the whole minor intestine increasingly, and operative in prognosticate gastrointestinal endoscopy. This will be attractive to patients and physicians specifically for cancer or ulcer detection, as capsule endoscopy is the pain-free check of the small intestine compared to conventional colonoscopy and gastroscopy. WCE is a pill-like device that consists of some tiny constituents. After a WCE is swallowed by a patient who has a diet for 12 hours, this tiny device begins to operate and record images while moving along the digestive tract. Meantime, the images recorded by the camera are sent out wirelessly to a particular receiver attached to a belt. This process lingers for about 8 hours until the WCE battery runs out. Eventually, all the images kept in the recorder are copied into a computer. The images should be coded into video if necessary, and finally, physicians can view the images or the video and analyze different reasons for diseases in the GI tract [1]. This product is confirmed by the U.S. Food and Drug Administration (FDA) [2]. However, due to the power concerns and hardware resource limitations [3], the image quality of capsule endoscopy is lower than that of conventional ways. Furthermore, WCE images suffer from blurriness, variance in the brightness of different areas, and high intensity in the areas close to the camera. Therefore, it is vital to enhance the quality of the WCE images for accurate feature extraction and classification[4]. Image enhancement aims to improve the visibility of the region of interest to the purpose that we have [5], and suppress noise [6].

In this paper, a novel method for WCE image quality improvement is proposed. The proposed technique consists of three significant steps. First, as a nonlinear intensity transformation, Adaptive Fraction Gamma Transformation (AFGT)[7] is applied to intensity values. AFGT can adaptively improve illumination. Afterward, an Unsharp Masking filter (UM) is applied to boost the details and textures. Finally, the color of the images is enhanced based on the improved intensity of the images. This method can effectively magnify luminance, contrast, and details, and also preserve the color information for WCE images. It is computationally simple, which is a potential algorithm for real-time applications.

The rest of the paper is designed as follows. In section II, some related works are discussed. Then, in Section III, we present the proposed algorithm included AFGT, UM filtering, and color restoration. Section IV presented the results based on the simulation results of MATLAB. In conclusion, Section V closes the paper.

## II. Related works

Image enhancement techniques can be divided into four categories: Non-linear Intensity Transformation (NIT) [8-10], Histogram Equalization (HE) [11-13], Unsharp Masking [14], and Retinex theory [15]. HE is one of the most straightforward algorithms for automatic contrast image adjustment, which is a classical approach for image enhancement[16,17]. The HE-based method is prevalent in image enhancement for its simplicity and suitable for images with no need for extensive modification [18]. On the other hand, this method causes annoying artifacts and unnatural appearance and magnifies noise. Furthermore, it can sometimes saturate the image and worsen the results for feature extraction. Moreover, for RGB images, if histogram equalization is applied to an RGB image, it will change the properties of the color channel (R, G, B), which results in changing the color of the image. Thus, the RGB image must be changed to LAB color space, and then HE is applicable. Under this condition, the hue and saturation of the image will not change. CLAHE seems more convenient to adjust the contrast [19], which is a modified version of Adaptive Histogram Equalization. The main modification between Adaptive Histogram Equalization (AHE), and Contrast Limited Adaptive Histogram Equalization (CLAHE) is contrast limiting [12]. The CLAHE provides a clipping limit for the histogram to overcome the noise problems. CLAHE tends to reduce the over enhancement of noise in the homogeneous regions of an image. Small areas in images are called a tile. CLAHE algorithm operates for each tile in the image, thus allows for more precise control of contrast. Brightness Preserving Dynamic Fuzzy Histogram Equalization (BPDFHE) is a novel modification algorithm for brightness while preserving the dynamic histogram equalization. BPDFHE algorithm improves image contrast also preserves the illumination very efficiently. Although this technique reduces the computational complexity, like other histogram-based approaches, in images with very dark or very bright areas, this algorithm will fail [13].

Nonlinear Intensity Transformation (NIT) algorithms as the second category of image enhancement methods enhance image contrast by mapping the intensity value of each pixel to another amount. These methods are simple and suitable for real-time applications [10,20].

UM algorithms concentrate on detail enhancement by processing high-frequency and low-frequency components distinctly. Which low-frequency components represent texture [21]. These algorithms generate artifacts, and usually, the color of enhanced images is not natural [14]. Retinex-based methods such as Robust Retinex [15] enhance images effectively and have better human visualization, but usually generate halo effects; besides, these algorithms are complex and time-consuming.

As explained the benefits and weaknesses of each method, it can be concluded that the choice of the image enhancement algorithm influenced totally by the specification of the capsule endoscopy image. For example, the UM methods focus on detail enhancement, and histogram-based methods focus on illumination adjustment. In the next section, we will analyze the details of the proposed algorithm.

### III. Proposed method

The block diagram of the proposed method is shown in Fig. 1. As given in Fig. 1, the proposed technique consists of three major steps: AFGT, Unsharp Masking filter, and color restoration. First, the RGB color image is reformed to HSI color space. This color space is a general model in machine vision applications. Hue, Saturation, and Intensity, can be achieved through simple transformations of the RGB space. Hue identifies the base color, saturation identifies the purity of a color, and intensity indicates how bright the color should be.

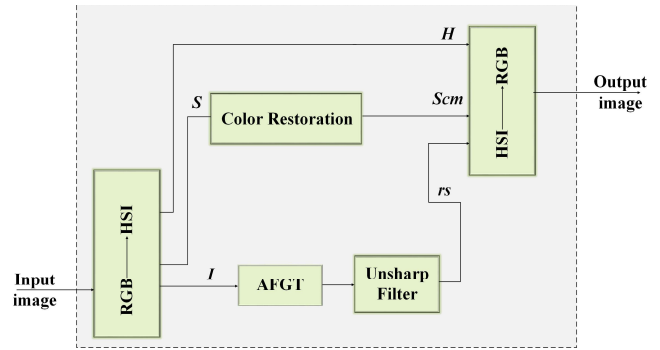

Fig. 1. The block diagram of the proposed method.

HSI space is much less sensitive to changes in brightness since HSI color space separates the intensity from the color information (hue and saturation) in a color image. Color changing due to variations in brightness is a typical problem of the WCE images as the battery of the capsule decreases over time. Using HSI space, we can handle this difficulty. Then we normalize the intensity component by Eq. (1):

$$I_n(i,j) = \frac{I(i,j)}{I_{max}} \quad (1)$$

here, $I_{max}$ is the maximum value of the intensity component, and $I_n(i,j)$ is the normalized intensity component. After intensity normalization, we will take advantage of AFGT as a nonlinear intensity transformation filter.

#### A. AFGT

We will apply the AFGT to the normalized intensity component. In [7], AFGT is defined as Eq. (2):

$$L(i,j) = \left(\frac{I_n^\gamma(i,j)}{2 - I_n^\gamma(i,j)}\right)^\beta \quad (2)$$

Where $L(i,j)$ is the modified intensity component. $\beta$ and $\gamma$ are adaptive parameters which are discussed below.

Gamma correction ($I_n^\gamma$) is broadly used to improve contrast. As shown in Fig. 2(a), when $\gamma > 1$, the large values of intensity are enhanced, while when $\gamma < 1$, the small values of intensity are amended. Since WCE images consist of both dark and light areas, $\gamma$ must change depending on the intensity values for both areas; therefore, $\gamma$ is defined as the Eq.(3):

$$\gamma = 1 + arctan(I_n(i,j) - 0.5) \quad (3)$$

As shown in Fig. 2 (b), if $I_n(i,j) > 0.5$, the function is an increasing function, where it tends to stretch large values of intensity, while for $I_n(i,j) < 0.5$, the small values of intensity are enhanced. For the determination of parameter $\gamma$, which is adaptive to different images and variables with different input intensities, the authors in [7] utilized the cumulative distribution function (cdf) of an image. For cdf calculation, the probability density function (pdf) is operational [22]. pdf is defined as Eq.

(4). Where $r_N$ is the intensity level, M and N are the size of the image.

$$pdf(r_N) = \frac{\sum\sum \delta(I_n, r_N)}{MN} \quad (4)$$

$$\delta(x,y) = \begin{cases} 1 & if\ x = y \\ 0 & else \end{cases} \quad (5)$$

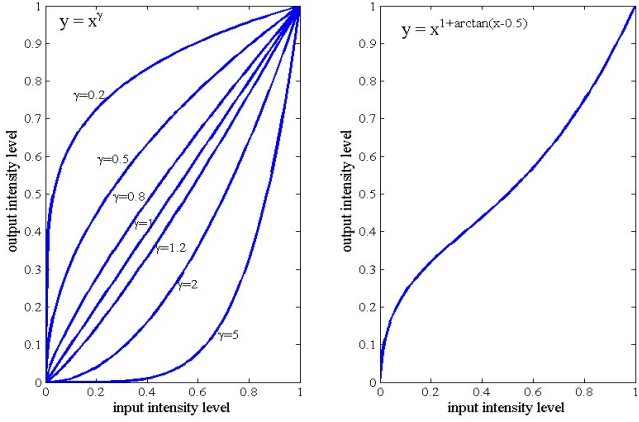

Fig. 2. Transformation curves. (a) Gamma correction. (b) Gamma correction with adaptive parameter

We define smoothed pdf as Eq. (6):

$$pdf_s(r_N) = pdf_{max}\left(\frac{pdf(r_N) - pdf_{min}}{pdf_{max} - pdf_{min}}\right)^\tau \quad (6)$$

here, $pdf_{max}$ is the maximum pdf value of the original image, $pdf_{min}$ is the minimum pdf value of the original image, $pdf_s$ is the smoothed pdf, and $\tau$ is the adjusted parameter. For the determination of parameter $\tau$, the standard deviation (std) of an image is beneficial:

$$\tau = I_{max}\left(\sum_{r_N=0}^{1}(r_N - \bar{r})^2 \cdot pdf(r_N)\right)^{\frac{1}{2}} \times 1\% \quad (7)$$

$$\bar{r} = \sum_{r_N=0}^{1} r_N \cdot pdf(r_N) \quad (8)$$

Now, the modified cdf is defined as the following function:

$$cdf_s(r_N) = \frac{\sum_{k=0}^{r_N} pdf_s(k)}{\sum_{r_N=0}^{1} pdf_s(r_N)} \quad (9)$$

here, $cdf_s$ is the modified cdf. And finally, the parameter β is defined as Eq. (10):

$$\beta(r_N) = \frac{1}{1 + cdf_s(r_N)} \quad (10)$$

Under calculating the presented parameters for each input image, the AFGT adaptively enhanced detail and contrast based on pixel intensity in an image and the standard deviation of different images. After AFGT filtering, we will apply an unsharpen mask filter.

### B. Unsharp mask filter

The high-frequency components mostly provide proper details of an image [23]. Therefore, they are essential and useful in many image processing applications [24-25]. Unsharp Masking (UM) is a classical approach for sharpness enhancement that uses High Pass Filter (HPF) to reach this goal [9]. This technique cannot create additional details, but it can significantly enhance the appearance of the details and emphasize the texture of an image by sharpening the image. Despite the simplicity and functional outcome in many applications, this method is sensitive to noise, and as mentioned before, WCE images due to the hardware limitation and the environment of the GI tract, are noisy. While in the definition of UM it uses HPF, but in HPF, noise passes and is not filtered out, therefore in this work, in order to extract high-frequency details from an image while removing the noise, we utilize the subtraction of normalized intensity component and the normalized intensity component that passing through Low Pass Filter (LPF) instead of high-pass filter block in the unsharp mask. It will reduce the vulnerability to noise. After that, this value is multiplied by a factor. However, in this work, after repeating the simulations several times considering various WCE images, we have obtained that the best threshold factor for this application is 0.8. The UM filter we proposed is defined as follows:

$$r_S(i,j) = L + 0.8 \times (I_n - LPF(I_n)) \quad (11)$$

where $L$ is the intensity that enhanced by AFGT in Eq. (2), and $I_n$ is the normalized intensity component of the input image. Here we use the Gaussian filter as the low pass filter. The block diagram of the UM filter is available in Fig. 3. After image enhancement steps, we will restore the color of the input image as the last step of the proposed algorithm.

### C. color restoration

To restore the color of enhanced intensity component, a linear color restoration is employed as following Eq. (12)[7]. After scaling, the dynamic range of $S_c$ is expanded. This is similar to contrast stretching. In [7], $S_c$ is mapped to [0,1] by Eq. (13). Where $sc_{max}$ and $sc_{min}$ are the maximum and minimum value $of\ S_c(i,j)$, relatively. In this approach outliers cause problems. A few pixels are at extreme values and Eq. (13) could lead to very unrepresentative scaling and reduced all pixel values, hence image is converted to gray scale image.

$$S_c(i,j) = S_i(i,j) \times \frac{r_s(i,j)}{I_n(i,j)} \quad (12)$$

$$S_{C,m}(i,j) = \frac{S_c(i,j) - sc_{min}}{sc_{max} - sc_{min}} \quad (13)$$

A more robust approach is achievable by first taking histogram of $S_c$, then clipping top and lower 0.2% of pixel values, and at the end, histogram equalization is used to map $S_c$ to [0,1]. The steps of the proposed scheme can be seen visually in Fig.4

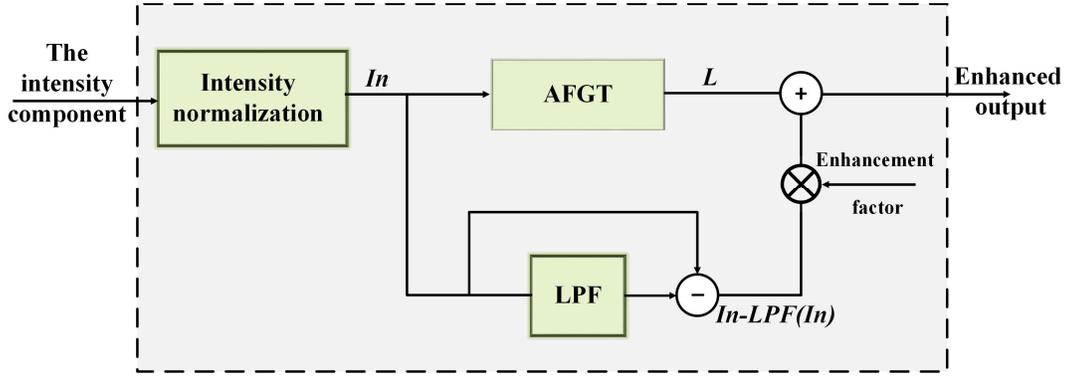

Fig. 3. The block diagram of UM filter.

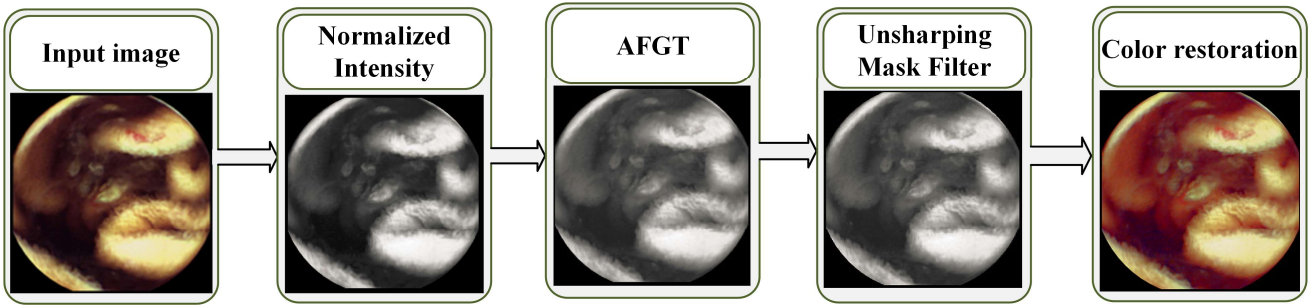

Fig. 4. Visual steps of the proposed method

## IV. Results

In this section, we evaluate the proficiency of the proposed technique. 300 various RGB images are randomly selected from the KID dataset [26]. The resolution of the images is 360×360. All implementations are performed on MATLAB R2014a. Image enhancement algorithms which prepared to compare are Robust Retinex [15], AGCWD [10], CLAHE[11], BPDFHE[13].

### A. Visual quality

Results enhanced by different methods are shown in Fig. 5, and Fig. 6 for two original images consist of both dark and light areas. As we see, the BPDFHE cannot reveal details marked red boxes and tend to generate artifacts in the bright area marked by blue boxes in Fig. 5(b) And Fig. 6(b), so its corresponding histograms have uncording distribution lying at a high-intensity level. The AGCWD method improves the luminance and the contrast, but it cannot reveal details in very dark areas marked by red boxes in Fig. 5(c) and Fig. 6(c), so the corresponding histograms have many pixels that lie in low-level intensities. The CLAHE algorithm generates obviously artifacts marked by blue boxes, and the color tends to be black-marked by green boxes in Fig. 5(d) and Fig. 6(d), hence the corresponding histogram has many intensity values lying in low-intensity levels. Also, CLAHE cannot announce details in dark areas. In Fig. 5(e) and Fig. 6(e), the Robust Retinex scheme reveals details in shady areas, but it distorts the lightness of the image. As shown in Fig. 5(f) and Fig. 6(f), the proposed method can announce details in dark areas, and enhance the color and illumination of the image as well. In the next section, we will compare the results from the objective assessment point of view.

### B. Objective Assessments

Metrics, including intensity restricted average local entropy (IRMLE)[7] based on Eq. (14), color enhancement factor (CEF) [27], lightness-order-error (LOE)[28], Peak-Signal-Noise-Ratio(PSNR), Structural Similarity Index Measure (SSIM) [29] are chosen to assess image quality objectively. MLE is commonly used to evaluate information richness. Since the WCE images have limited contrast, MLE and entropy are not exact metrics for WCE image. IRMLE is calculated by Eq. (14). In this equation, m and n are the sizes of the image, and $LE(i,j)$ is local entropy of a 9×9 window centered at pixel $(i,j)$. IRMLE depicts the information richness of an image. The large values depict more details. We used LOE to objectively measure the lightness distortion of the enhanced images. The smaller LOE value indicates better naturalness preservation. We used CEF to observe the quality in terms of color enhancement; the high values of CEF represents better color quality. Average results of 300 images for different methods are listed in Table I. CLAHE has the best IRMLE value, but it has lousy LOE value, and it cannot protect the natural color of images, but as mentioned before because of the simplicity of this algorithm, its suitable for real-time applications.

$$IRMLE = \frac{1}{mn}\sum_{i=1}^{m}\sum_{j=1}^{n}LE(i,j)\sum_{r_N=1/3}^{2/3}pdf(r_N) \qquad (14)$$

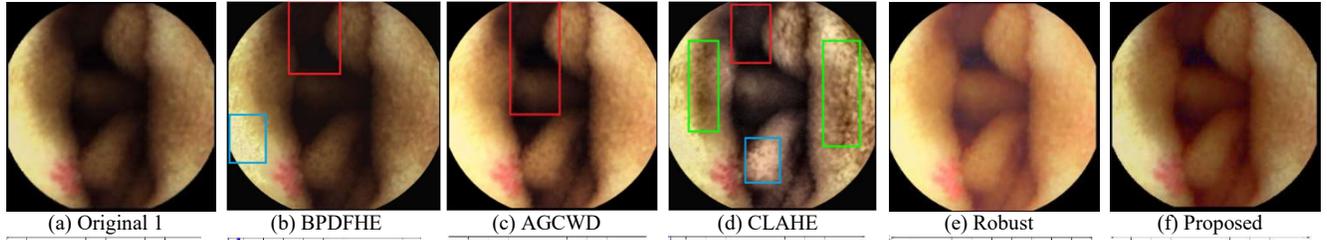
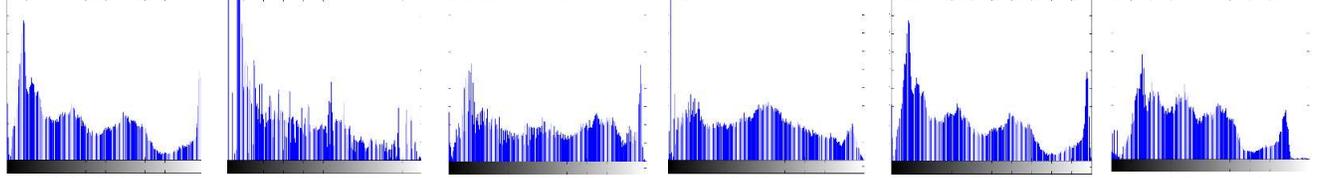

Fig.5. Comparisons of results enhanced by different methods. (a) Original image 2. (b) Enhanced by BPDFHE. (c) Enhanced by AGCWD. (d) Enhanced by CLAHE. (e) Enhanced by Robust. (f) Enhanced by proposed method.

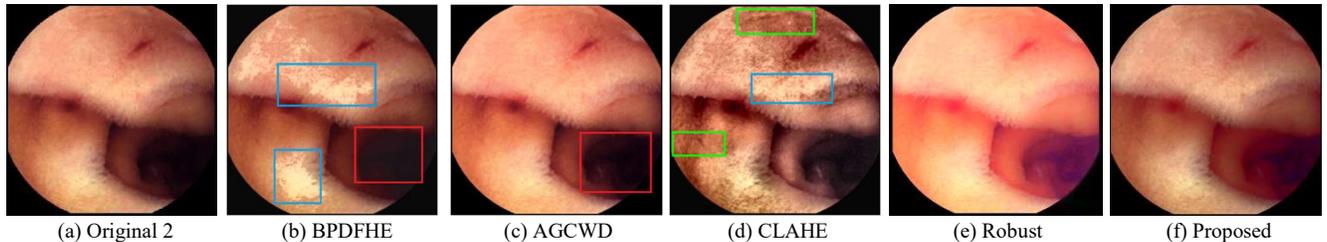
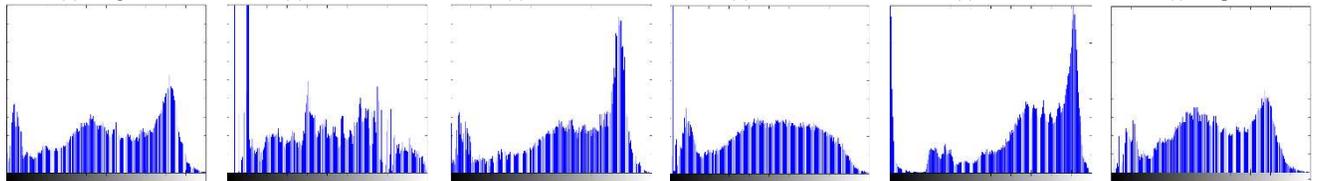

Fig. 6. Comparisons of results enhanced by different methods. (a) Original image 2. (b) Enhanced by BPDFHE. (c) Enhanced by AGCWD. (d) Enhanced by CLAHE. (e) Enhanced by Robust. (f) Enhanced by proposed method.

BPDFHE has better LOE value, but it has lower IRMLE value even compared to the original images, this value shows that BPDFHE cannot reveal details and cannot enhance illumination. This algorithm can be used for real-time applications. AGCWD has good CEF and SSIM values, but its IRMLE value is lower even in comparison with the original images, so AGCWD is not suitable for WCE image enhancement. This algorithm is also suitable for real-time applications. Robust Retinex has unfavorable results for improving WCE images and not ideal for these images; otherwise, this algorithm is not appropriate for real-time applications. Fig.7 (a), (b), (c), (d), and (e) represent the CEF, IRMLE, LOE, PSNR, SSIM for each testing image for all of the methods, respectively. As shown in Table I and Fig. 6, the proposed algorithm has been more successful in eliminating noise than the other techniques and presents desirable IRMLE, CEF, LOE, and SSIM values. Experimental results demonstrate that the proposed methods can enhance illumination with qualified details and preserves color information as well.

## V. Conclusion

In this research, we presented a new method for image enhancement on low contrast and low-quality WCE images. The proposed algorithm focuses on enhancing the contrast and quality of the capsule endoscopy images. The experimental results of SSIM, PSNR, IRMLE, LOE, and CEF demonstrate that the proposed method can enhance the illumination and details of the capsule-based endoscopy images effectively. With the idea of AFGT and the proposed way for unsharpen mask filter, noise in the images is reduced efficiently. While with the color restoration algorithm, we could keep the color information at the same time. These results also show that our proposed method has outstanding performance compared with state-of-the-art methods. Our proposed method is applicable to many other applications, such as microscopy image enhancement, surveillance image enhancement, and biometric image enhancement.

Table I. Performance evaluation of image enhancement algorithms

| Methods | ori. | CLAHE | BPDFHE | AGCWD | Robust Retinex | Pro. |
|---|---|---|---|---|---|---|
| IRMLE | 1 | 1.28 | 0.97 | 0.85 | 0.62 | 1.22 |
| CEF | 1 | 1 | 1.01 | 1.17 | 0.007 | 1.1 |
| LOE | 0 | 4.63 | 0.21 | 0.24 | 5.5 | 0.69 |
| PSNR | - | 21.19 | 26.74 | 28.80 | 8.1 | 28.91 |
| SSIM | 1 | 0.71 | 0.72 | 0.96 | 0.28 | 0.95 |
| Real time | - | Yes | Yes | Yes | No | Yes |

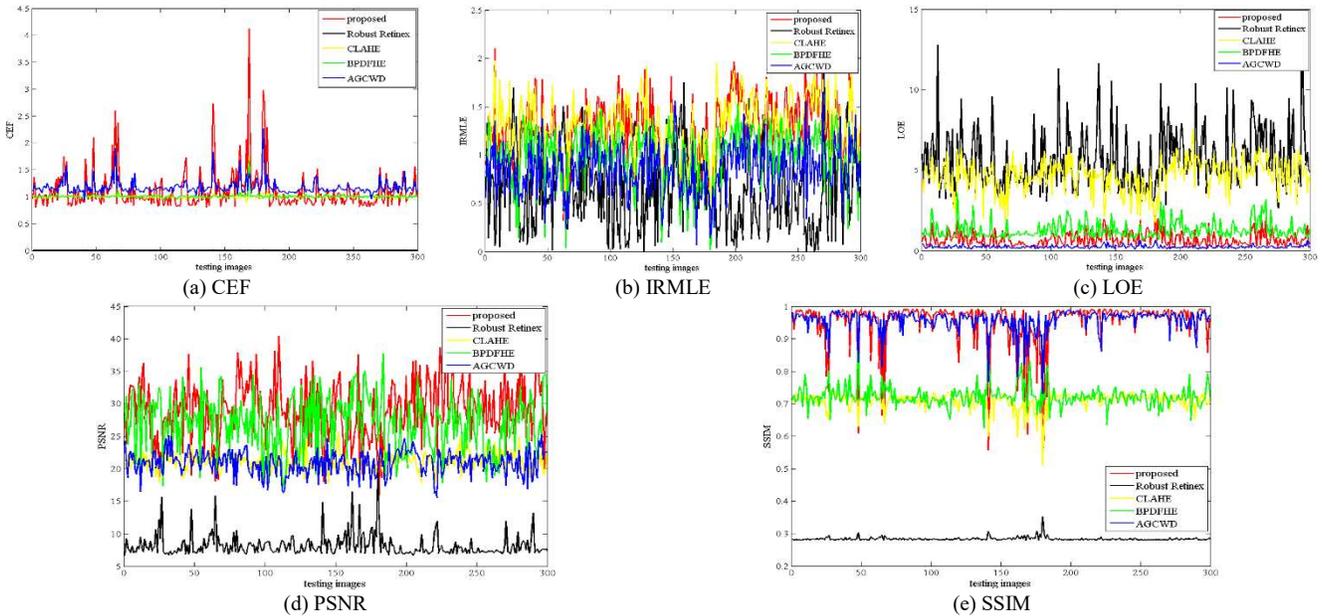

(a) CEF  (b) IRMLE  (c) LOE  (d) PSNR  (e) SSIM

Fig. 7. Objective assessments for the 300 testing images. (a) CEF, the larger the better. (b) IRMLE, the larger the better, (c) LOE, the smaller the better, (d) PSNR, the larger better, (e) SSIM, the larger better

References

[1] A. Hoffman, C. Kagel, M. Goetz, A. Tresch, J. Mudter, S. Biesterfeld, P. R. Galle, M. F. Neurath, and R. Kiesslich,."Recognition and characterization of small colonic neoplasia with high-definition colonoscopy using i-Scan is as precise as chromoendoscopy," *Digestive and Liver Disease.*, vol. 42, no. 1, pp. 45–50, 2010.

[2] P. Swain, "The future of wireless capsule endoscopy," World J. Gastroenterol., vol. 14, no. 26, pp. 4142–4145, 2008.

[3] B. Li and M. Q. H. Meng, "Wireless capsule endoscopy images enhancement via adaptive contrast diffusion," J. Vis. Commun. Image Represent., vol. 23, no. 1, pp. 222–228, 2012.

[4] M. Long, Z. Li, X. Xie, G. Li, and Z. Wang, "Adaptive Image Enhancement Based on Guide Image and Fraction-Power Transformation for Wireless Capsule Endoscopy," IEEE Trans. Biomed. Circuits Syst., vol. 12, no. 5, pp. 993–1003, 2018.

[5] M. Heidari, A. Z. Khuzani, A. B. Hollingsworth, et al. "Prediction of breast cancer risk using a machine learning approach embedded with a locality preserving projection algorithm." Physics in Medicine & Biology, vol.63, no. 3, p.035020, 2018.

[6] M. Heidari, S. Samavi, S.M.R. Soroushmehr, et al. "Framework for robust blind image watermarking based on classification of attacks," Multimed Tools Appl, vol.76, no.22, pp.23459–23479, 2017.

[7] M. Long, Z. Lan, X. Xie, G. Li, and Z. Wang, "Image Enhancement Method Based on Adaptive Fraction Gamma Transformation and Color Restoration for Wireless Capsule Endoscopy," 2018 IEEE Biomed. Circuits Syst. Conf. BioCAS 2018 - Proc., pp. 1–4, 2018.

[8] G. Zahi and S. Yue, "Adaptive intensity transformation for preserving and recovering details in low light images," Proc. Comput. Conf. 2017, pp. 262–271, 2017.

[9] P. Sidike, V. Sagan, M. Qumsiyeh, M. Maimaitijiang, A. Essa, and V. Asari, "Adaptive Trigonometric Transformation Function with Image Contrast and Color Enhancement: Application to Unmanned Aerial System Imagery," IEEE Geosci. Remote Sens. Lett., vol. 15, no. 3, pp. 404–408, 2018.

[10] S. C. Huang, F. C. Cheng, and Y. S. Chiu, "Efficient contrast enhancement using adaptive gamma correction with weighting distribution," IEEE Trans. Image Process., vol. 22, no. 3, pp. 1032–1041, 2013.

[11] M. Moradi, A. Falahati, A. Shahbahrami, and R. Zare-Hassanpour, "Improving visual quality in wireless capsule endoscopy images with contrast-limited adaptive histogram equalization," 2015 2nd Int. Conf. Pattern Recognit. Image Anal. IPRIA 2015, no. Ipria, pp. 0–4, 2015.

[12] A. M. Reza, "Realization of the contrast limited adaptive histogram equalization (CLAHE) for real-time image enhancement," J. VLSI Signal Process. Syst. Signal Image. Video Technol., vol. 38, no. 1, pp. 35–44, 2004.

[13] D. Sheet, H. Garud, A. Suveer, M. Mahadevappa, and J. Chatterjee, "Brightness preserving dynamic fuzzy histogram equalization," IEEE Trans. Consum. Electron., vol. 56, no. 4, pp. 2475–2480, 2010.

[14] G. Deng, "A generalized unsharp masking algorithm," IEEE Trans. Image Process., vol. 20, no. 5, pp. 1249–1261, 2011.

[15] M. Li, J. Liu, W. Yang, X. Sun, and Z. Guo, "Structure-Revealing Low-Light Image Enhancement Via Robust Retinex Model," IEEE Trans. Image Process., vol. 27, no. 6, pp. 2828–2841, 2018.

[16] D. Iakovidis, "Software Engineering Applications in Gastroenterology," Glob. J. Gastroenterol. Hepatol., vol. 2, no. 1, pp. 11–18, 2014.

[17] M. Heidari, S. Mirniaharikandehei, A. Z. Khuzani, et al. "Improving performance of CNN to predict likelihood of COVID-19 using chest X-ray images with preprocessing algorithms." arXiv preprint arXiv:2006.12229, 2020.

[18] A.Z. Khuzani, M. Heidari, S. A. Shariati, "COVID-Classifier: An automated machine learning model to assist in the diagnosis of COVID-

19 infection in chest x-ray images." medRxiv, 2020.
[19] K.Zuiderveld, "Contrast limited adaptive histogram equalization," Graph. Gems IV. San Diego Acad. Press Prof., pp. 474–485, 1995
[20] M. Heidari, A. Z. Khuzani, G. Danala, et al. "Improving performance of breast cancer risk prediction using a new CAD-based region segmentation scheme." Medical Imaging, vol. 10575, p. 105750P, 2018.
[21] M. Heidari, N. Karimi, S. Samavi. "A hybrid DCT-SVD based image watermarking algorithm." Iranian Conference on Electrical Engineering (ICEE), pp. 838-843, 2016.
[22] M. Heidari, S. Mirniaharikandehei, W. Liu, et al. "Development and assessment of a new global mammographic image feature analysis scheme to predict likelihood of malignant cases." IEEE Transactions on Medical Imaging, vol.39, no. 4, pp.1235-1244, 2019.
[23] A. Z. Khuzani, G. Danala, M. Heidari, et al. "Applying a new unequally weighted feature fusion method to improve CAD performance of classifying breast lesions." Medical Imaging, vol. 10575, p. 105752L, 2018.
[24] M. Heidari, S. Gaemmaghami. "Universal image steganalysis using singular values of DCT coefficients." 10th International ISC Conference on Information Security and Cryptology (ISCISC), pp. 1-5. IEEE, 2013.
[25] A. Zargari, Y. Du , et al. "Prediction of chemotherapy response in ovarian cancer patients using a new clustered quantitative image marker." Physics in Medicine & Biology, vol.63, no. 15, p.155020, 2018.
[26] D. Koulaouzidis and A. Iakovidis, "KID: A capsule endoscopy database for medical decision support," United Eur. Gastroenterol. Week, 2015.
[27] S. E. Susstrunk and S. Winkler, "Color Image Quality on the Internet," Internet Imaging V, vol. 5304, pp. 118–131, 2003.
[28] S. Wang, J. Zheng, H. M. Hu, and B. Li, "Naturalness preserved enhancement algorithm for non-uniform illumination images," IEEE Trans. Image Process., vol. 22, no. 9, pp. 3538–3548, 2013.
[29] Z. Wang, A. C. Bovik, H. R. Sheikh, and E. P. Simoncelli, "Image quality assessment: from error visibility to structural similarity," IEEE Trans. Image Process., vol. 13, no. 4, pp. 600–612, 2004.